\documentclass{ws-p8-50x6-00}

\usepackage{subfigure}

\def\Asymm{\mathcal{A}}
\def\Dalitz{D}
\def\etal {{et al.,}\ }
\begin{document}

\title{New results from NA48 experiment
 on neutral kaon rare decays}

\author{A. Bizzeti}

\address{Universit\`a di Modena e Reggio Emilia,
 Dipartimento di Fisica,
\\ via G. Campi, 213/A,
 I-41100
 Modena, Italy
\\
 INFN - Sezione di Firenze,
 largo E. Fermi 2,
 I-50125
 Firenze, Italy
}

\address{{\rm
 representing the NA48 Collaboration: \\
Cagliari, Cambridge, CERN, Dubna,
 Edinburgh, Ferrara, Firenze,
\\
 Mainz, Orsay, Perugia, Pisa, Saclay,
 Siegen, Torino, Vienna, Warsaw
}}

\maketitle

\abstracts{
Recent results by the NA48 collaboration
 on
 neutral kaons
 rare decays
into the
 $\pi^+\pi^-e^+e^-$
 final state
are presented.
A large CP-violating asymmetry
 $ \Asymm (K_L) = ( 13.9 \pm 2.7 \pm 2.0 ) \% $
 has been observed
 in the $K_L \to \pi^+\pi^-e^+e^-$ decay.
The $K_S \to \pi^+\pi^-e^+e^-$ decay
has been observed for the first time,
 showing no such asymmetry.
}

\section{Introduction}\label{sec:intro}

The
 NA48 experiment at CERN SPS,
 described in detail elsewhere\cite{epsilon,tim},
has been designed to measure
 the direct CP violation parameter
 $ \Re (\epsilon'/\epsilon) $
 in two-pion decays of neutral kaons.
However, the quality of its simultaneous
 $K_L$ and $K_S$ beams\cite{beam},
 high resolution detectors,
 fast trigger
 and performing
  data acquisition system
allow to investigate
  at the same time
 rare decays of neutral kaons
  and neutral hyperons.
Rare decay data have been collected during
 1997-1999
 $ \Re (\epsilon'/\epsilon) $ runs,
as well as during
 dedicated runs 
 with a high intensity $K_S$ and hyperon beam
 in 1999 and 2000.

In this talk I will
 report recent results from the NA48 experiment
 on $K_L$ and $K_S$ decays into the
  $\pi^+ \pi^- e^+ e^-$ final state,
which
 provide a new opportunity
  to probe CP violation in the neutral kaon sector.

\section{$K_L \to \pi^+ \pi^- e^+ e^-$}\label{sec:kl_pipiee}

The $e^+e^-$ pair in
 $\pi^+ \pi^- e^+ e^-$
 decays of neutral kaons
 is expected to originate from a virtual photon\cite{sehgal,heiliger}:
 $K_{L,S} \to \pi^+ \pi^- \gamma^* \to \pi^+ \pi^- e^+ e^-$.
The $K_L \to \pi^+\pi^- e^+e^-$ decay amplitude has two
 main
 components,
 one from CP-odd
  magnetic dipole (M1)
  direct photon emission,
 the other from the decay to the
 CP-even state $\pi^+ \pi^-$
  with inner bremsstrahlung.
Interference
 between these
 amplitudes
 causes a large
 CP-violating
 asymmetry
 in the distribution of the
  angle $\varphi$ between the normals
   to the $\pi^+\pi^-$ and $e^+e^-$ planes
   in the kaon c.m. system:
\begin{equation}
{{\rm d}\Gamma \over {\rm d}\varphi} = \Gamma_1 \cos^2(\varphi)
 + \Gamma_2 \sin^2(\varphi )
 + \Gamma_3 \sin(\varphi)\cos(\varphi)
   \  .
\label{eq:pipiee_dgamma_dphi}
\end{equation}
A non-zero value of $\Gamma_3$,
 i.e. a dependence of d$\Gamma$/d$\varphi$
on the sign of the CP-odd and T-odd variable
  $sin(\varphi)cos(\varphi)$,
 constitutes a clear signature of CP violation
 which can be seen in the d$\Gamma$/d$\varphi$ distribution.
The CP-violating asymmetry
\begin{equation}
\Asymm = {
       N_{ \sin(\varphi)\cos(\varphi) > 0 } 
     - N_{ \sin(\varphi)\cos(\varphi) < 0 }
    \over
       N_{ \sin(\varphi)\cos(\varphi) > 0 }
     + N_{ \sin(\varphi)\cos(\varphi) < 0 }
}
\label{eq:pipiee_asymm}
\end{equation}
 is predicted by theory\cite{sehgal,heiliger}
 to be
$ | \Asymm ( K_L
 ) | = 15\% \cdot sin ( \phi_{+-} + \delta_0(M^2_K) - \bar\delta_1 )
   \approx 14\% $.
The observation of such a large asymmetry has been reported
 recently\cite{ktev}
 and found
  to be
  in agreement
 with theoretical predictions.

During 1998 and 1999 run periods
 more than 1300 good $ K_L \to \pi^+ \pi^- e^+ e^- $ events
 have been selected
using a dedicated four-track trigger\cite{ks_pipiee}.
The event identification
 relies
 on the precise tracking by the magnetic spectrometer
and on the good energy resolution of the
 electromagnetic calorimeter
for $e/\pi$ separation.
Background from
 $K_L \to \pi^+ \pi^- \pi^0_{\Dalitz} \to \pi^+ \pi^- e^+ e^- \gamma$
 decays is
 strongly suppressed
 using
 a kinematic cut\cite{p02},
%
while
 $\pi^+\pi^- e^+e^-$ events
 from two
 overlaid
 $K_{e3}$ decays
  are rejected by time constraints.
Electrons from photon conversions
  in the material
 before
 the spectrometer
 are eliminated
by requiring a 2~cm minimum separation
 between the $e^+$ and $e^-$ tracks in the first chamber.

\begin{figure}[b]
\begin{center}
\epsfxsize=0.8\textwidth
\epsfbox{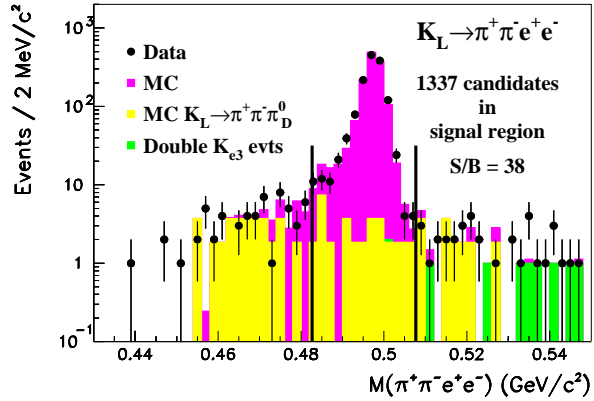} 
\caption{Invariant mass
 distribution of $K_L \to \pi^+ \pi^- e^+ e^-$ candidate events.
\label{fig:kl_pipiee_signal}}
\end{center}
\end{figure}

Figure \ref{fig:kl_pipiee_signal} shows the invariant mass distribution
 of the $K_L \to \pi^+ \pi^- e^+ e^-$ candidate
 events passing all analysis cuts,
together with the expected background contributions.
\begin{figure}[b]
\begin{center}
\epsfxsize=10pc
\epsfxsize=0.8\textwidth
\epsfbox{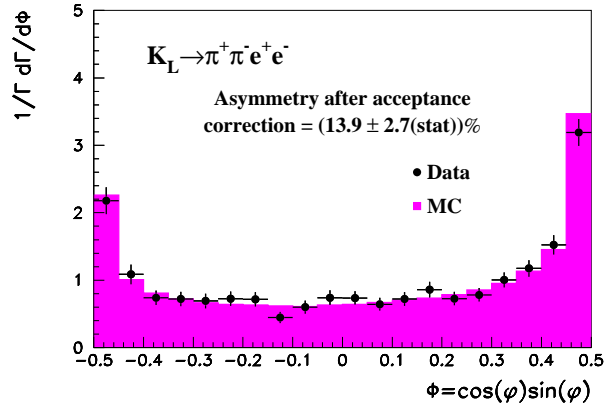} 
\caption{Acceptance corrected $sin(\varphi) cos(\varphi)$ distribution
 of $K_L \to \pi^+ \pi^- e^+ e^-$ candidate events.
\label{fig:kl_pipiee_asymmetry}}
\end{center}
\end{figure}
For the determination of the $K_L \to \pi^+ \pi^- e^+ e^-$
 branching fraction
fully reconstructed
 $K_L \to \pi^+ \pi^- \pi^0_{\Dalitz} \to \pi^+ \pi^- e^+ e^- \gamma$
 decays 
are used as normalization channel.
The correction for the detector acceptance is
 determined using a model by Heiliger and Sehgal\cite{heiliger}
 with the inclusion of a form factor in the M1 direct emission amplitude:
\begin{equation}
F_{M1} =
{\tilde{g}_{M1} }
 \biggl[ 1 + {
{a_1/a_2}
 \over (M^2_\rho - M^2_K) c^2 + 2 M_K
 ( E_{e^+}+E_{e^-} )
 } \biggr]
\label{eq:fm1}
\end{equation}
Using
 the values
 $\tilde g_{M1} = 1.35^{+0.20}_{-0.17} $
 and
 $a_1/a_2 = ( -0.720 \pm 0.029 ) $ (GeV/c)$^2$
 measured by KTeV\cite{ktev}
 we obtain the preliminary result
$ B ( K_L\to\pi^+\pi^-e^+e^- )
 = ( 3.1\pm 0.1 \pm 0.2 ) \cdot 10^{-7} $,
in fair agreement with the KTeV preliminary result\cite{senyo}
of $ ( 3.63 \pm 0.11\pm 0.14 ) \cdot 10^{-7}$.

Figure \ref{fig:kl_pipiee_asymmetry} shows the
 $sin(\varphi)cos(\varphi)$
 distribution of
 $ K_L \to \pi^+ \pi^- e^+ e^- $ events
after acceptance correction.
Our preliminary result on the observed asymmetry
 is  $ \Asymm (K_L) = ( 13.9 \pm 2.7 \pm 2.0 ) \% $,
in good agreement with theoretical
 predictions\cite{heiliger}
and the recently published KTeV value\cite{ktev}.


\section{$K_S \to \pi^+ \pi^- e^+ e^-$}\label{sec:ks}

The
  $K_S \to \pi^+ \pi^- e^+ e^-$ decay amplitude
is largely dominated by
 the CP-even inner bremsstrahlung process,
 so no CP-violating asymmetry
 is expected in this case.

This decay mode has been observed for the first time
 with a clean sample of 56 events
 selected from 1998 data\cite{ks_pipiee}.
The decay events coming from the $K_S$ beam
 have been selected
 using the very good vertex resolution
  of the magnetic spectrometer
 and
 the precise timing of the tagging detectors\cite{tagger}.
\begin{figure}[b]
\begin{center}
\epsfxsize=0.8\textwidth
\epsfbox{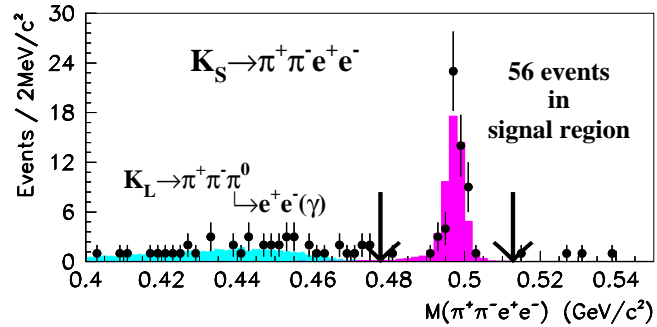} 
\caption{
Invariant mass
 distribution
 of $K_S \to \pi^+ \pi^- e^+ e^-$ candidate events (1998 data).
\label{fig:ks_pipiee_1998}}
\end{center}
\end{figure}
The invariant mass distribution of
 $K_S \to \pi^+ \pi^- e^+ e^-$ candidate events
is shown in figure \ref{fig:ks_pipiee_1998}.
Low invariant mass events in the plot are due to
 $K_L \to \pi^+ \pi^- \pi^0_{\Dalitz} \to \pi^+ \pi^- e^+ e^- \gamma$
 decays of $K_L$ mesons originating from the $K_S$ target.
Using 105 fully
 reconstructed $K_L\to\pi^+\pi^-\pi^0_{\Dalitz}$
 events from the $K_S$ target
 as normalization channel
 from 1998 data
 a branching fraction
 $B(K_S \to \pi^+ \pi^- e^+ e^-)
  = ( 4.5 \pm 0.7 \pm 0.4 ) \cdot 10^{-5}$.

In 1999 two days of run
 were dedicated to the investigation
  of rare decays of $K_S$ and neutral hyperons,
with the proton beam intensity on the $K_S$ target
 increased by about a factor 200,
the $K_L$ beam switched off
and the AKS converter at the beginning
 of the fiducial volume removed.
The large amount of rare decay events collected
 during this short test run,
equivalent to several years of operation
 with the standard $K_L+K_S$ beam setup,
 allowed a more precise determination of the branching fraction
 and a measurement of the CP-violating asymmetry $\Asymm (K_S)$
 at a few percent level.
\begin{figure}[b]
\begin{center}
\epsfxsize=0.8\textwidth
\epsfbox{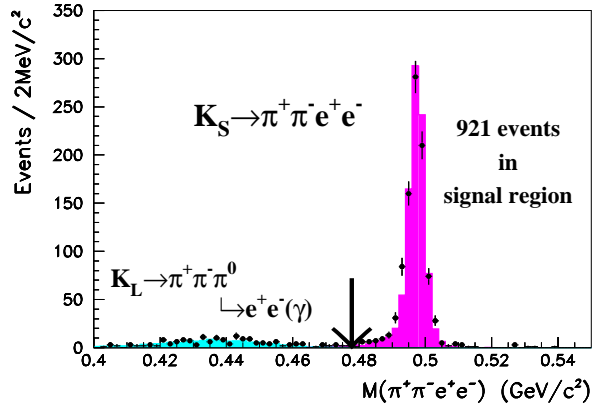}   
\caption{Invariant mass
 distribution of $K_S \to \pi^+ \pi^- e^+ e^-$ candidate events
 (1998+1999 data).
\label{fig:ks_pipiee_signal_1999}}
\end{center}
\end{figure}
\begin{figure}[b]
\begin{center}
\epsfxsize=0.8\textwidth
\epsfbox{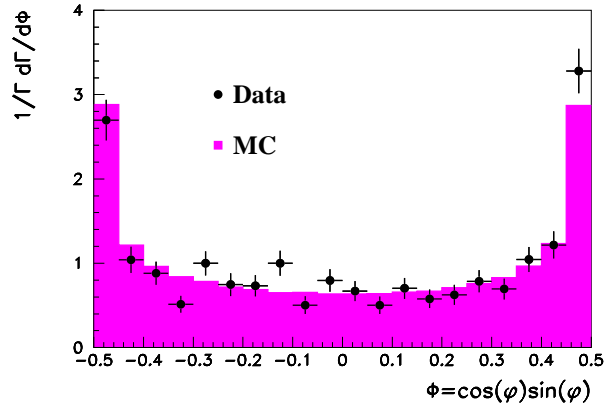} 
\caption{Acceptance corrected $sin(\varphi)cos(\varphi)$ distribution
 of $K_S \to \pi^+ \pi^- e^+ e^-$ candidate events
 (1998+1999 data).
\label{fig:ks_pipiee_asymm_1999}}
\end{center}
\end{figure}
The invariant mass
 distribution
of $K_S \to \pi^+ \pi^- e^+ e^-$ candidate events
 from the full 1998+1999 data sample
 is shown in figure \ref{fig:ks_pipiee_signal_1999}.
The
 $sin(\varphi)cos(\varphi)$ distribution
  of these events
  is plotted in figure \ref{fig:ks_pipiee_asymm_1999}:
the resulting asymmetry
 $\Asymm (K_S) = ( -0.2 \pm 3.4 \pm 1.4 ) \% $
is consistent with zero.
Our
 preliminary
 result on the branching fraction is
$ B(K_S \to \pi^+\pi^- e^+ e^-)
 = (4.3 \pm 0.2 \pm 0.3 )
 \cdot 10^{-5} $.
Using this number to evaluate
 the inner bremsstrahlung contribution
  to the
 $K_L$
  decay we obtain
 $ B(K^{IB}_L \to \pi^+\pi^- e^+ e^-)
  = (1.3\pm 0.1) \cdot 10^{-7}$,
in good agreement with theoretical
 expectations\cite{sehgal,heiliger}.

\end{document}